\documentclass[epj]{webofc}
\usepackage[varg]{txfonts}   
\usepackage[T1]{fontenc}
\usepackage[utf8]{inputenc}
\usepackage{amsmath}
\usepackage{amssymb}
\usepackage{graphicx}
\usepackage{esint}
\woctitle{The Time Machine Factory 2012}
\begin{document}
\title{Wormholes and Time-Machines in Nonminimally Coupled Matter-Curvature Theories of Gravity}
%
%

\author{Orfeu Bertolami\inst{1}\fnsep\thanks{\email{orfeu.bertolami@fc.up.pt}}\and
		Ricardo Zambujal Ferreira\inst{2}\fnsep\thanks{\email{ferreira@cp3.dias.sdu.dk}}$^{,}$\thanks{Speaker}
}

\institute{Departamento de Física e Astronomia, Faculdade de Ciências, Universidade do Porto, Rua do Campo Alegre 687, 4169-007 Porto, Portugal
\and
CP3-Origins \& Danish Institute for Advanced Study DIAS, University of Southern Denmark, Campusvej 55, DK-5230 Odense M, Denmark
          }

\abstract{In this work we show the existence of traversable wormhole and time-machine solutions in a modified theory of gravity where matter and curvature are nonminimally coupled. Those solutions present a nontrivial redshift function and exist even in the presence of ordinary matter which satisfies the dominant energy condition.
}
\maketitle
\section{Introduction}
\label{intro}
General Relativity (GR) is one of the great pillars of modern physics
and it is based on the idea that the curvature of space-time
is dynamically tied up with the matter distribution. GR has been well
tested in most of its foundations \cite{Bertolami:2006js} and it accounts for all gravitational phenomena locally, and also globally, if one admits dark components. Indeed, GR can account for the cosmological observations only if one assumes that the Universe is predominantly dominated composed by two
yet unknown constituents: dark energy $\left(\Omega_{DE}\simeq73\%\right)$
and dark matter $\left(\Omega_{DM}\simeq23\%\right)$. This implies
that only $4\%$ of the content of the Universe is baryonic.

Despite its extraordinary agreement with the observations, it is believed that
GR cannot be regarded as the final theory (see e.g. Ref. \cite{Bertolami:2011aa} for a survey).

The first modifications to GR were suggested by Weyl,
in 1919 \cite{Weyl}, and Eddington, in 1923 \cite{Eddington}. However, the
motivation for modifying GR became more acute when it was realized that it
is not compatible with Quantum Mechanics.

At this point there are two posible alternatives.
One can consider that there are really new and unknown dark constituents
dominating the Universe or one assumes that the observed features with only ordinary matter are deviations
from GR \cite{Bertolami:2006fh}.

A prototype model of the first alternative is the standard $\Lambda$CDM that assumes
the existence of a cosmological constant to account for dark energy
and weak-interacting particles, arising from extensions to
the Standard Model, to account for dark matter.

The second approach, which will be adopted in this
contribution, consists in assuming that the need of invisible components to account for the observations, actually signal the need to extend GR. There are various ways to generalize
GR.  In this work we will consider an extension of the so-called $f(R)$ theories \cite{RevModPhys.82.451}.

\section{Nonminimally coupled matter-curvature theories of gravity}
\label{NMC} 
In $f(R)$ theories one replaces the scalar curvature $R$
in the gravitational action by an arbitrary function of $R$. There
are two main motivations for doing so. First, this is the simplest
way of modifying the action and, second, this replacement can account
for the main features of higher order theories of gravity. However,
our attention will be focused on an extension of $f(R)$ theories
which has received a great deal of attention recently. In this extension matter
and curvature are nonminimally coupled, that is NMC modified theories of
gravity.

This approach does possess some new and interesting
features. In NMC theories, besides the change in the gravitational
sector like in $f(R)$ theories, one also couples the Lagrangian density of matter
nonminimally with the curvature by an arbitrary function of the Ricci
scalar \cite{Bertolami:2007gv}:

\begin{equation}
S=\int\left[\frac{1}{2k}f_{1}\left(R\right)+\left(1+\lambda f_{2}\left(R\right)\right){\cal {\cal L_{M}}}\right]\sqrt{-g}d^{4}x.\label{eq:Effective stress-energy tensor definition}
\end{equation}

NMC theories have striking properties such as the nonconservation
of the energy momentum tensor, which leads to deviation from geodesic
motion of test particles \cite{Bertolami:2007gv}, putative
deviations from hydrodynamic equilibrium of stars \cite{Bertolami:2007vu},
and the breaking of the degeneracy of the Lagrangian densities, which
in GR give rise to the energy-momentum tensor of perfect fluids \cite{Bertolami:2008ab}.

Moreover, these models allow for the mimicking of dark
matter profiles in galaxies \cite{Bertolami:2009ic}
and clusters of galaxies \cite{Bertolami:2011ye}, as well as of dark
energy at cosmological scales \cite{Bertolami:2010cw}.
It was also shown that one can obtain somewhat more natural conditions
for preheating in inflationary models \cite{Bertolami:2010ke} and
that the coupling between curvature and matter is equivalent, under
conditions, to an effective pressure which leads, in the weak field
limit, to a generalized Newtonian potential \cite{Bertolami:2011rb}. It has also been shown that NMC between curvature and matter can mimic, for a suitable matter distribution, a cosmological
constant \cite{Bertolami:2011fz}.

In this work we review the results of Ref. \cite{PhysRevD.85.104050}, where it was investigated the possibility
of generating closed timelike curves (CTCs) in this specific modified
theory of gravity. 

\section{Wormholes and Time-Machines}
\label{TM} 
A CTC is a closed worldline in space-time. This apparently simple
feature has disturbing consequences as such a curve can globally violate causality \cite{Bertolami:2009er}. A related issue concerns the fact that microscopic systems
are described by laws which are invariant under time reversal. This
means that one can start with some initial conditions and evolve forward
in time or, equivalently, start from future conditions and evolve
the system towards the past. Nevertheless, this feature is not shared by the macroscopic world as the second law of thermodynamics
provides a well defined time direction. 

However, the possibility of curving time in GR implies that causality
might be violated. In fact, several CTCs solutions are known
in GR, the first one discovered by van Stockum in 1937 \cite{Stockum}.
One can ask whether CTCs can be avoided by any physical mechanism
or if CTCs are possible without paradoxes. Currently, there is no complete
and satisfactory answer to this issue and this has led to conjectures
such as the well known Chronology Protection Conjecture of Stephen
Hawking \cite{H} (see the contribution of Orfeu Bertolami in this volume for a more detailed discussion).

Nevertheless, if one allows for time-travel or exchange of
information between the past and future, one has to
ensure some properties besides the creation of CTCs. In Ref.
\cite{Morris:1988cz} a detailed analysis of the conditions to allow
for an interstellar journey of beings through a traversable wormhole
is presented and in a subsequent work (Ref. \cite{Morris:1988tu})
it was shown how to convert traversable wormholes into time-machines.
It is known that such conversion is not unique \cite{Visser:1992tx}.

In GR, in order to construct time-machine solutions, exotic
types of matter, which violate energy conditions such as
the Null Energy Condition (NEC), $T_{\mu\nu}k^{\mu}k^{\nu}<0$, are
required. Furthermore, a quantum analysis of
matter configurations under these conditions show that CTCs become unstable
due to several effects like the Casimir effect, gravitational back-reaction
and others \cite{Visser:1992tx}. Hence, time-travel turns out to
be most likely impossible.

In the context of NMC theories, wormhole geometries have been studied
for the particular case of trivial red-shift functions \cite{Garcia:2010xb,MontelongoGarcia:2010xd},
the function that defines the $g_{00}$ component of the metric. In
the work of Ref. \cite{PhysRevD.85.104050} these results were extended and connected with the possibility
of time-travel through traversable wormholes. The main issue
in NMC theories is that the NEC violation close to the wormhole throat concerns the effective energy-momentum tensor (c.f. Ref. \cite{Bertolami:2009cd}), that is $T_{\mu\nu}^{eff}k^{\mu}k^{\nu}<0$, where and $k^{\mu}$ is a null vector which, for simplicity, is chosen to be radial $k^{\mu}=k^{0}\left(1,\sqrt{\frac{-g_{00}}{g_{rr}}},0,0\right)$ and
\begin{equation}
T_{\mu\nu}^{eff}=\frac{1}{F_{1}}\left[\left(\Delta_{\mu\nu}-\frac{R}{2}g_{\mu\nu}\right)F_{1}(R)+\frac{1}{2}g_{\mu\nu}f_{1}(R)+2\lambda\left(\Delta_{\mu\nu}-R_{\mu\nu}\right){\cal L_{M}}F_{2}(R)+\left(1+\lambda f_{2}(R)\right)T_{\mu\nu}^{(m)}\right],\label{eq:Stress Energy Tensor}
\end{equation}
with $\Delta_{\mu\nu}=\nabla_{\mu}\nabla_{\nu}-g_{\mu\nu}\square$.

Moreover, using results from the literature, we demonstrate
that time-machines can be created, under conditions, from wormholes in static spherically symmetric space-times described by the metric,
\begin{equation}
ds^{2}=-e^{2\Phi(r)}dt^{2}+\frac{dr^{2}}{1-\frac{b(r)}{r}}+r^{2}\left(d\theta^{2}+\sin^{2}\theta d\phi^{2}\right),\label{eq:spherically symmetric space-time metric}
\end{equation}
where $\Phi\left(r\right)$ and $b\left(r\right)$ are arbitrary functions,
usually referred to as redshift function and shape function, respectively.

\section{Results}
\label{R} 

Thus, in the context of NMC modified theories of gravity, we sought
for those solutions in the simplest case of a nonminimal coupling
where $f(R)=R$ \cite{PhysRevD.85.104050}. The field equations
were then solved for a perfect fluid threading the wormhole with isotropic
pressure and two different energy densities. One constant and localized
given by $\rho\left(r\right)=\rho_{0}\Theta\left(r-r_{2}\right)$,
where $r_{2}$ is some arbitrary scale, and another exponentially decaying
given by $\rho_{2}(r)=\frac{\rho_{0}r_{0}}{r}e^{-\frac{r-r_{0}}{\sqrt{2\lambda}}}$,
where $r_{0}$ is the wormhole throat. For these energy densities,
the shape function was obtained. Then, the system of equations was
solved for the redshift function and the pressure in two limits, $r\rightarrow r_{0}$
and $r\rightarrow\infty$, in order to ensure the violation of the
NEC and a suitable asymptotic behavior.

In the first case, wormhole solutions were obtained with a constraint
on the coupling parameter: $\lambda>1/\left(2\rho_{0}\right)$. Moreover,
these solutions can be obtained even for ordinary matter if $\rho_{0}>\frac{1}{2\lambda}\left(1+\frac{r_{0}}{\sqrt{2\lambda+r_{0}^{2}}}\right)$.
However, the localized and constant energy density $\rho_{1}\left(r\right)$
has an intrinsic problem as it implies an unavoidable discontinuity
at an arbitrary scale. Hence, although wormhole solutions are found,
these are not traversable since the discontinuity leads to unphysical
regimes.

Concerning the second case, for the energy density $\rho_{2}\left(r\right)$,
wormhole solutions were obtained for a physical region of parameters $\left(\rho_{0},r_{0},\lambda\right)$. Furthermore, it was also found a region in the parameter space where these solutions are well defined
and satisfy the Dominant Energy Condition for ordinary matter $\left(\rho>0\right)$. In
this case, discontinuities, horizons and ill defined functions are
avoided. Moreover, since the solution is obtained for ordinary matter,
it is stable.

\section{Conclusions}
\label{C} 

Thus, we can conclude that well behaved CTCs and wormholes solutions
are posible in NMC models of gravity. Furthermore, these solutions can
be transformed in traversable wormholes which allow for time-travel of ordinary matter
if some quantitative conditions, both for the wormhole and for the
acceleration which yields the time shift, are satisfied.

It is relevant to stress that, as expected, the two wormhole solutions
obtained do not include the case $\lambda=0$, which corresponds to
GR. Thus, the crucial point in generating the wormhole solutions lies
in the presence of this nonminimal coupling between the curvature
and the matter.

%
%
%

\end{document}